\documentstyle[12pt]{article}
\vspace{5cm}
\title{Perturbation theory in the nonperturbative QCD vacuum}
\vspace{2cm}
\author{Yu.A.Simonov\\ Institute of Theoretical and Experimental Physics\\
117259, Moscow}
\date{}
\begin{document}
\maketitle
\vspace{1cm}

\begin{abstract}

Perturbative expansion in the nonperturbative confining QCD background is
formulated. The properly renormalized $\alpha_S(R)$ is shown to be finite at
large distances, with the string tension playing the role of an infrared
regulator. The infra--red renormalons are shown to be absent in the new
perturbation series.
\end{abstract}

\section{Introduction}

The perturbation theory (PTH) expansion in QCD is an extremely useful tool,
which has provided numerous links between theory and experiment [1].

The physical foundation of PTH in QCD is the celebrated asymptotic
freedom phenomenon [2], ensuring the smallness of the physical
coupling constant $\alpha_s$ at large momenta (small distances).

A formal extrapolation of the lowest order (one-loop) expressions
 for the renormalized $\alpha_s$ to large distances (small momenta)
is known to be plagued by the ghost pole, which signals an uncontrolled growth
of the strength of interaction at large distances. It is believed that PTH
expansion breaks down even before the Landau ghost pole is reached,
so that the latter has no physical meaning, similarly to the case
of QED [3,4].

The impossibility of treating large distances
in PTH of QCD strongly limits applications of QCD and from the
formal point of view signifies the incompleteness of the theory.

The logarithmic growth of the renormalized $\alpha_s$ at large distances
leads also to a basic difficulty, which is known as the infrared (IR)
renormalon problem [5]. The IR renormalons appear as singularities on the
positive
real axis of the Borel parameter and destroy an attempt to sum up the
PHT series.

Thus the consistency of the PHT approach is seriously in doubt, when
one expands around the free unperturbed vacuum.

Alternatively it was  understood during the last
decades, that nonperturbative (NP) contributions are very important at large
distances. In the framework of the QCD sum rules [6] the NP contributions haven
been parametrized as local field vacuum averages using the operator
product expansion (OPE) [7].

The phenomenologically found characteristics of the QCD vacuum clearly
demonstrate
 that the NP fields in the vacuum are large (as compared to $\Lambda_{QCD}$).
E.g. the gluonic condensate $<G^aG^a>\approx (600 MeV)^4\gg \Lambda_{QCD}^4$
[6]; lattice calculations yield even larger values [8].

The NP effects coded as power corrections in OPE [6] crucially change the
behavior of the OPE series (as compared to the PTH series) when one is
leaving the asymptotic freedom  region and approaching large distances. To
quote the seminal paper [6]:  "The basic idea  behind all applications is
that it is the power terms (not higher orders in the $\alpha_s$ series) that
limit asymptotic  freedom, if one tries to extend the short-distance
approach to larger distances".

It is specific to the {\underline local} OPE, that power terms explode at
large distances, demonstrating the importance of the NP effects but also the
difficulty of the method to penetrate the region beyond the confinement radius.

In this paper we propose to make a step forward in our strategy and instead
of expansion in local condensates to consider from
the beginning the PTH  in the NP background of a general form. The framework
for
this exists for a long time [9-11].
Gluonic field is split into the background $B_{\mu}$ and perturbative gluon
field
$a_{\mu}$ and the new PTH is roughly an  expansion in powers of $g~a_{\mu}$.
The propagators of $a_{\mu}$ are exact Green's functions $G_{\mu\nu}$ in the
background field $B_{\mu}$.

A new element, crucial for the method, is that we are using a closed form for
$G_{\mu\nu}$
based on the so-called Feynman-Schwinger representation (FSR) [12]. This
allows to display explicit dependence on $B_{\mu}$, which enters
through the parallel transporters $P~exp~ig\int^x_y B_{\mu}dz_{\mu}$.  In
the limit of large $N_c$ any term of the new perturbative series contains
$B_{\mu}$ through the Wilson loops $W(C,B)$.  Averaging over an ensemble of
background fields $\{ B_{\mu} \}$ then reduces the graph to a product of
averaged Wilson loops if $N_c \rightarrow \infty $, and all the
nonperturbative series is unambigiously defined by the expression for the
averaged Wilson loop $<W(C,B)>$. At large distances the latter obeys the
area law [13], and one obtains therefore a one-parameter description of NP
dynamics in the new perturbation series.

As  a result one obtains a gauge-invariant PTH, even though the
background gauge fixing term is used in the Lagrangian together with
Faddeev-Popov
ghosts [14]. It is not a purpose of this paper to formulate Feynman rules
for  diagrams of the new PTH and to study systematically the renormalization
procedure in the case of the NP background. Instead we concentrate here on the
lowest one-loop contribution to the charge renormalization. In this way
we attempt to answer the question: how the asymptotic freedom
(AF) behaviour of $\alpha_s$, which is generated by the lowest gluon and ghost
corrections in the background field, is maintained and how background
parameters enter into the final expression for $\alpha_s$.

To do that one first must define the charge renormalization
in a gauge-invariant and NP nonlocal formulation. We propose 3
definitions and two of them study quantitatively.

We find that the NP background fields change the renormalized charge in a
drastic way for small momenta (large distances).

The usual logarithmic growth of $\alpha_s(R)$ at large distances in the
one-loop
approximation disappears and $\alpha_s(R)$ saturates instead at rather
moderate values. The mass scale $m$ (or distance $m^{-1}$) which governs
 this change of asymptotics is large, $m\sim 1 Gev$, and physically is
connected to the hybrid mass -- or in the string terms -- to the excitation
mass
of the transverse string vibration, $ m^2 \sim 2\pi\sigma \sim 1 GeV^2$.

As a result  the asymptotic freedom can be kept at any distances in the
Euclidean space and expresses itself as a saturation of $\alpha_s$
at $\alpha_s(max) \leq 0.5$.

We always use Eucledian metrics and do not analytically continue
perturbative expansions into the Minkowskian region, where physical
thresholds appears. This will be a topic of separate publication.

This property drastically changes the problem of the summation of the PTH
series.
We show that the IR renormalons do not appear in the background field,
and the corresponding set of graphs has a finite sum.

Hence the only obstacle left is instantons (instanton - antiinstanton pairs) ,
which require an explicit treatment .

We stress that it not the purpose of our method to suggest
an explicit way of splitting the fields into the background and perturbative
ones.
Instead we choose from the beginning the NP background fields,
 they are considered as given and  this input is
parametrized, e.g. by the full set of correlators
$<F(x_1)F(x_2)...F(x_n)>$.

The problem one has to solve in this case is how to find the perturbative
series in
this background, and we attempt to do it in the paper.
Happily it appears that at large distances the only characteristics of the
background
which enters the answer is the string tension, and we take it to have the
known phenomenological value.

The plan of the paper is as follows In section 2 we introduce background field
Lagrangian, gauge-fixing and ghost terms following the usual prescriptions
[9-11]
and write explicitly the gluon propagator.
In section 3 three definitions of the running coupling are
compared, and in Section 4 one of it -- defined
through the energy of static quarks, is studied in detail. Another
definition, yielding the running coupling $\alpha_s(p)$ in the momentum
 space, is discussed in Section 5.

Section 6 is devoted to the renormalons and the problem of summation of the
perturbative series.

In the conclusion we summarize results of the paper anl outline
possible future developments.

Three appendices devoted to the explicit derivations of formulas used or
discussed
 in the text. In Appendix 1 we consider the contribution of the term $L_1$,
which directly connects NP background and the perturbative field $a_{\mu}$,
to the propagator of $A_{\mu}$.

Appendix 2 contains some detailes of derivation of the renormalization
$\alpha_s(R)$ using the Wilson-loop approach.

 Appendix 3 contains a
detailed disscussion of the one-loop graph in presence of the confining
background; most of the results used in the main text are based on the
formulas  explicitly derived there.

 In Appendix 4 we treat the gluon spin part of the gluon propagator in the
 background field. Finnaly, Appendix 5 is devoted to the explicit
 calculation of the IR renormalon contribution.

\section{Background Lagrangian and propagators}

We split the gluonic field $A_{\mu}$ into the background $B_{\mu}$ and
perturbative part $a_{\mu}$
\begin{equation}
A_{\mu}= B_{\mu}+a_{\mu}
\end{equation}

As we stressed in the Introduction we do not specify any principle of this
division.
For the purposes of the paper the exact properties of  separation are
immaterial, and we shall only assume that the NP fields $B_{\mu}$
ensure the area low of  the Wilson loop $<W(B)>$
with roughly the same string tension $\sigma$ as known phenomenologically,
$\sigma \approx 0.2 GeV$. To develope the PTH in the background field $B_{\mu}$
we use the well established background field formalism [9-11], specifically we
extensively use method and notations of the paper [11], adjusted to the
Euclidean
space-time.

The generating functional can be written as

\begin{equation}
Z(J) = \int Da ~det(\frac{\delta G^a}{\delta w^b}) exp  \int d^4 x
[L(a)-\frac{1}{2}(G^a)^2 + J^a_{\mu} a^a_{\mu}]
\end{equation}
where $L(a) = L_0+L_1(a)+L_2(a)+L_{int}(a)$

\begin{eqnarray}
L_2(a)&=&+\frac{1}{2} a_{\nu}(\hat{D}^2_{\lambda}\delta_{\mu\nu} -
\hat{D}_{\mu}\hat{D}_{\nu} + ig \hat{F}_{\mu\nu}) a_{\mu}= \nonumber \\
&=&\frac{1}{2} a^c_{\nu}[D_{\lambda}^{ca}D_{\lambda}^{ad} \delta_{\mu\nu}
- D_{\mu}^{ca}D_{\nu}^{ad} - g~f^{cad}F^a_{\mu\nu}]a^d_{\mu}~~,
\end{eqnarray}
\begin{eqnarray}
D_{\lambda}^{ca} &=& \partial_{\lambda}\cdot \delta_{ca}+ g~f^{cba}
B^b_{\lambda}
\equiv \hat{D}_{\lambda}
\nonumber
\\
L_0 &=& -\frac{1}{4} (F^a_{\mu \nu}(B))^2 ~;~~~ L_1=a^c_{\nu} D_{\mu}^{ca}(B)
F^a_{\mu\nu}
\nonumber
\\
L_{int} &=& -\frac{1}{2} (D_{\mu}(B)a_{\nu} -D_{\nu}(B)a_{\mu})^a
g~f^{abc} a_{\mu}^b a_{\nu}^c - \frac{1}{4} g^2 f^{abc} a_{\mu}^b a_{\nu}^c
f^{aef} a^e_{\mu} a^f_{\nu}
\end{eqnarray}

The background gauge condition is

\begin{eqnarray}
G^a = \partial_{\mu}a^a_{\mu} + g f^{abc} B^b_{\mu} a^c_{\mu} =
(D_{\mu} a_{\mu})^a
\end{eqnarray}
and the ghost vertex [14] is obtained from $\frac{\delta G^a}{\delta \omega^b}
= (D_{\mu}(B)D_{\mu} (B+a))_{ab}$ to be
\begin{eqnarray}
L_{ghost}=-\theta^+_a (D_{\mu}(B)D_{\mu} (B+a))_{ab} \theta_b
\end{eqnarray}

The  linear part part of the  Lagrangian $L_1$ disappears
if $ B_{\mu} $ satisfies classical equations of motion. We  do not
impose on $B_{\mu}$  this condition, and we show in the Appendix 1 that
$L_1$ gives no contribution to the effect we are interested first of all --
 the modification of the AF logarithm.

  We now can identify the propagator of $a_{\mu}$ from the quadratic terms
in Lagrangian $L_2(a) - \frac{1}{2\xi}(G^a)^2$
\begin{eqnarray}
G^{ab}_{\nu\mu} = [\hat{D}^2_{\lambda} \delta_{\mu \nu} -
\hat{D}_{\mu}\hat{D}_{\nu} + ig \hat{F}_{\mu\nu} +\frac{1}{\xi}
\hat{D}_{\nu} \hat{D}_{\mu}]^{-1}_{ab}
\end{eqnarray}
It will be convenient sometimes to choose $\xi =1$
and end up with the well-known form of propagator in -- what one would call --
the background Feynman gauge
\begin{eqnarray}
G^{ab}_{\nu\mu} = (\hat{D}^2_{\lambda}\cdot  \delta_{\mu \nu} -
2ig \hat{F}_{\mu\nu})
^{-1}
\end{eqnarray}

\section{Definitions of the running coupling}

 In case of the free vacuum one can use different schemes of regularizations
e.g.
$MS$ or ${\overline MS}$ schemes ) and renormalize the coupling $g^2$,
fixing the value of the vertex at some Euclidean momentum. For, example one may
require that the quark (or gluon) propagator has the form of the free
propagator at some  momentum $p$, $p^2=-\mu^2$ [15].

In the case of NP vacuum this kind of definition is not possible
anymore and one should instead introduce a new NP notion of the regularized
coupling.
We consider here three examples of such definition .\\

i) \underline{Definition of the running coupling through the interaction}\\
 \underline{energy of heavy quarks (HQ)}\\

The interaction energy of a heavy quark and antiquark $E(R)$ at a distance $R$
between them can be written as
\begin{eqnarray}
E(R) = E_{NP} (R) - \frac{4}{3} \frac{\bar{\alpha}_s(R)}{R}
\end{eqnarray}
where $E_{NP}$ is the purely $NP$ part of the energy, while
$\bar{\alpha}_s(R)$ is a renormalized running coupling at distance $R$
\begin{eqnarray}
\bar{\alpha}_s (R) = \frac{\bar{g}^2(R)}{4\pi} = \frac{g_0^2}{4\pi}
(1+ \frac{g_0^2}{4\pi} f(R) + ...
\end{eqnarray}
and $g_0$ is the bare coupling, $f(R)$ is  to be calculated from the 1-loop
diagrams in the background $NP$ field.

In case of free vacuum $\bar{\alpha}_s^{(0)}(R)$ has been computed
to one loop in [16] through the bare coupling
\begin{eqnarray}
g^2(R) = g^2_0 + g^4_0 \frac{11}{48\pi^2} C_2 \ln (\frac{R}{\delta})^2
\end{eqnarray}
where $\delta \sim 1/M$ is the $UV$ cut-off.

Renormalizing via $\overline { MS}$ scheme with the normalization mass
$\mu$ one gets [17]
\begin{eqnarray}
\alpha_{\bar{M}\bar{S}} (\mu,R) = \alpha_{\bar{M}\bar{S}} \{ 1 +
\frac{\alpha_s}{\pi} [ \frac{b_0}{2} (ln \mu R + \gamma_E) + \frac{5}{12}
b_0 - \frac{2}{3} C_A] \}
\end{eqnarray}
with $b_0 = \frac{11}{3} N_c - \frac{2}{3} N_f~,~~ \gamma_E = 0.5777...$

The main feature of (11) and (12) is the logarithmic grouth of $\alpha_s(R)$
 at large distances, and this feature produces both the Landau ghost
pole [3,4] and the IR renormalon problem [5].
In the next sections we shall demonstrate that the strong confining NP
background eliminates the logarithmic growth of  $\alpha_s(R)$    and is
this way solves both the ghost pole and IR renormalon problem.

 To calculate  $\bar {\alpha}_s(R)$  in the NP background we shall use the
gauge-invariant
definition
\begin{eqnarray}
E(R) = -\lim_{T\rightarrow \infty} \{ \frac{1}{T} ln <W_C(B+a)>^{reg}_{B,a}
\}
\end{eqnarray}
where contour $C$ of the Wilson loop $<W_C>$ is a
rectangular $R\times T$ shown in Fig. 1.

The averaging over fields $B_{\mu}, a_{\mu}$ and regularization is
assumed in (13) and will be specified in the next section.
Note that $E(R)$ (13) contains both PTH and NP
contributions as in (9). We again stress at this point, that in our approach
there is
no problem of separation of a given $E(R)$ into $E_{NP}$ and
a PTH piece (which is of course    not well defined). Instead we
assume that an ensemble of fields $\{B_{\mu}\}$ is given to us, it is
characterized
by correlators
\begin{eqnarray}
g^n<F_{\mu_1\nu_1}(B,x_1)... F_{\mu_n \nu_n} (B,x_2)>_B
\end{eqnarray}
and in particular by the string tension $\sigma$, which can be expressed
through
the correlators. Our task is to construct the PTH in this given background
and to express in particular  $\bar{\alpha}_s(R)$ through the
nonperturbative parameters.

To perform renormalization in the background field formalism, one may use
its gauge invariance [10]. One can visualize that all terms in (2) are gauge
invariant with respect to gauge transformations "rotating" both $B_{\mu}$
and $a_{\mu}$. It is convenient to split gauge transformations as follows
[11]
\begin{eqnarray}
B_{\mu} \rightarrow U^+ (B_{\mu} +\frac{i}{g}\partial_{\mu}) U~,
\nonumber
\\
a_{\mu} \rightarrow U^+ a_{\mu}U~,~~B_{\mu}=B^a_{\mu}+t^a.
\end{eqnarray}

If we consider following [11] only amplitudes with no external $a_{\mu}$
lines (and we need only those), then renormalization  of $a_{\mu}$ and
$\theta$ lines is inessential (corresponding Z factors compensate each
other), while $g$, $B, \xi$ renormalize as

\begin{eqnarray}
g_0= Z_gg~, ~ B_0 =Z^{1/2}_BB, ~~\xi_0 = Z' \xi
\end{eqnarray}
Now because of the explicit gauge invariance,
$Z_g$ and $Z_B$ are connected  [11]
\begin{eqnarray}
Z_g^{-1}~= ~Z^{1/2}_B
\end{eqnarray}
as a consequence, the quantities $gB_{\mu}$ and $gF_{\mu\nu}(B)$ are RG
invariant, and therefore all characteristics of NP fields, like
correlators (14) or string tention are RG invariant and should be
considered on the same footing as the external momenta in the amplitudes.\\

ii) \underline{Definition through the $B$-field two-point function}\\

 This is equivalent to the method used in [11], where two 1-loop diagrams
 have been used of Fig. 2 and 3 with two external $B$ lines. To make
 averaging over fields $\{B_{\mu}\}$ we multiply each diagram in the
 coordinate space with the parallel transporter in the adjoint
 representation,
\begin{eqnarray}
\hat{\Phi} (x,y;B) = P exp~ ig~ \int^{x}_y \hat{B}_{\mu}dz_{\mu}
\end{eqnarray}
and consider both gluon and ghost propagating in the field $B_{\mu}$ with
propagators $G^{ab}_{\mu\nu}(x,y,B)$ and $G^{ab}(xy;B)$ respectively.
The product
\begin{eqnarray}
H(x,y)
\equiv <\hat{\Phi}^{ef}(x,y,B)\Gamma_{\mu}^{eac}G^{ab}_{\mu\nu}(x,y,B)
G^{cd}(x,y,B)\Gamma_{\nu}^{fbd}>_B
\end{eqnarray}
depicted in Fig. 4, yields the one-loop correction to the renormalized
charge, which in absence of field $B_{\mu}$ reduces to the second term on
the r. h. s. of (11). Using results of the next section one can again prove
that dependence like $ln|x-y|$ disappears in (19) when $|x-y|\gg m^{-1}, m>1
GeV$. This will be clear from discussion in the next two sections.

In a similar way one calculates two- and more loop corrections, using the
same two-point diagrams as in [11].\\

iii) \underline{Definition through the finite box of size $L$.}\\

This definision is similar to that of refs. [18]. One fills the box $L^4$
with the $NP$ field $B_{\mu}$ and considers one-loop correction to $g^2$. As
in [18] one gets
\begin{eqnarray}
\bar{g}^2(L)=g^2_0 +n_1(L)g^4_0 + ...
\end{eqnarray}
with $n_1 = h_0 -\frac{1}{2}h_1$,
\begin{eqnarray}
h_0=\frac{\frac{\partial}{\partial\eta}ln
{}~det\bar{G}}{\frac{\partial}{\partial\eta}\Gamma_0(B)} \;\;,
h_1=\frac{\frac{\partial}{\partial\eta}ln
{}~det G}{\frac{\partial}{\partial\eta}\Gamma_0(B)}
\end{eqnarray}

Here $\eta$ is a dimensionless parameter characterising strength of field
$B_{\mu}, \Gamma_0(B)$ is the effective action $(=\frac{1}{4}\int
F^2_{\mu\nu}(B)d^4 x)$.

For the abelian (nonconfining) field used in [18] one obtains the
asymptotics [18]
\begin{equation}
n_1(L)\sim_{L\rightarrow\infty}\sum^{\infty}_{n=0}(r_n +s_n ln L)/L^n
\end{equation}
with $r_n, s_n$ some constants, $s_0=\frac{11}{12\pi^2}$. Thus one recovers
the usual free-vacuum renormalization with the infra-red logarithmic growth.
If instead one uses the confining field $B_{\mu}$, the asymptotics of
$n_1(L)$ changes and the logarithm $ln L$ disappears (the study of $ln det
G$ in case of confining background was done in [19] for the case of nonzero
temperature $T$, a similar analysis for $T=0$ will be published elsewhere).

In the next sections we shall study in detail the first and second
definitions of the running coupling and shall find its one-loop
renormalization. As will be clear both definitions yield the same
qualitative phenomenon: the running coupling $\alpha_s(R)$ saturates at
large $R$ at some finite value, and the mass parameter $m$, which cuts off
the logarithmic growth of $\alpha_s(R)$, is of the same order in both
definitions.

\section{Calculation of $\alpha_s(R)$ through the Wilson loop formalism.}

In this section we shall follow the way outlined in the first definition of
$\alpha_s(R)$, i.e. we shall use $E(R)$ given by (13). To this end we need
the PTH expansion of the Wilson loop $<W(B+a)>$, i.e. expansion of the
powers in gluon operator $a_{\mu}$. We start with the Wilson contour $C$ of
an arbitrary configuration and divide the contour into $N$ pieces $\Delta
x_{\mu}(n)$, writing identically
\begin{eqnarray}
W(B+a)&\equiv&\frac{1}{N_c}tr P exp ig\int_{C}(B_{\mu}+a_{\mu})dx_{\mu}=
\nonumber \\
&=& \frac{1}{N_c}tr P\prod^{N}_{n=1}[1+
ig(B_{\mu}(x(n))+a_{\mu}(x(n)))\Delta x_{\mu}(n)]= \\
&=& \frac{1}{N_c}tr P\left\{\prod^{N}_{n=1}(1+
ig B_{\mu}\Delta x_{\mu})+
\prod^{k-1}_{n=1}(1+ig B_{\mu}\Delta x_{\mu})ig a_{\mu}(k)\Delta
x_{\mu}(k)\right.\times \nonumber \\
\nonumber
&\times&\left.\prod^{N}_{n=k+1}(1+ig B_{\mu}\Delta x_{\mu}(n))+ ...\right\}
\end{eqnarray}

One can represent the sum (23) as a sum of  Wilson loops $W^{(n)}(B)$ with
$n$ insertions of the field $ig a_{\mu}(k)$ on the contour $C$:
\begin{equation}
W(B+a) = W(B)+\sum^{\infty}_{n=1}(ig)^{n}W^{(n)}(B;X(1) ...
X(n))dx_{\mu_1}(1) ... dx_{\mu_n} (n)
\end{equation}
Now we must integrate over the fields $D a_{\mu}$ with the weight given in
(2). In the lowest order $<W(B)>_{a}=W(B)$ for the properly normalized
weight, and in the order $O(g^2)$ one has a one-"gluon" exchange diagram
shown in Fig. 5.
\begin{eqnarray}
-g^2<W^{(2)}>_a dx dy =
-g^2\int\Phi^{\alpha\beta}(x,y;B)G_{\mu\nu}^{\delta\alpha;\beta\gamma}
(x,y;B)\Phi^{\gamma\delta}(y,x,B)dx_{\mu}dy_{\nu}
\end{eqnarray}
where $\Phi^{\alpha\beta}$ are parallel transporters (18) with fundamental
indices written explicitly, $\alpha, \beta =1,...N_c$
\begin{equation}
G_{\mu\nu}^{\delta\alpha,\beta\gamma}=t^{\delta\alpha}_a t^{\beta\gamma}_b
G_{\mu\nu}^{ab} \;\; ,
\end{equation}
and $G^{ab}_{\mu\nu}$ is the gluon propagator in the $NP$ background field
$B_{\mu}$ given in (7) and (8).

It is appropriate at this point to specify the notion of the gluon in the
background and to distinguish it from the free gluon, interacting only
perturbatively. We propose to call it the "hybrid-gluon", or $h$-gluon,
since as we shall see, in most cases it will aquire a mass and behaves as a
hybrid.

In the next order, $0(g^4)$, one obtains five diagrams shown in Figs. 6-10.

Next step is ti average over the ensemble of $NP$ fields $\{B_{\mu}\}$. It
is not necessary to specify the ensemble and prescribe the measure; we can
use the cluster expansion to obtain for the simplest term [13].
\begin{equation}
<W(B)>_B = exp\sum^{\infty}_{n=1}\frac{(ig)^n}{n!}\int d\sigma(1) ...
d\sigma(n)<F(1) ... F(n)>
\end{equation}

Exact gauge invariant form of the correlator $<F(1) ... F(n)>$ is discussed
in [13]. The lowest order term in (27) e.g. has two independent structures
the Kronecker one and the full derivative term [13]
\begin{equation}
g^2<F_{\mu\nu}(u)\Phi(u,v)F_{\rho\sigma}(v)\Phi(v,u)>=(\delta_{\mu\rho}\delta_{\nu\sigma}-
\delta_{\mu\sigma}\delta_{\nu\rho})D(u-v)+\partial(D_1)
\end{equation}

It is $D$ and the Kronecker structures of higher correlators that bring
about the area law of the Wilson loop [13]:
\begin{equation}
<W(B)>_B \cong exp(-\sigma S_{min})\;\; , \;\; S~~~ large
\end{equation}
where
\begin{equation}
\sigma =\frac{1}{2}\int\int^{\infty}_{-\infty}d^2 u D(|u|)+\int<FFFF>+ ...
\end{equation}

Thus any finite member of terms in the sum (27) or the whole sum, if it
converges, yields the area law for the Wilson loop. In what follows we shall
assume that the area law with $\sigma\cong\sigma_{phen}$ holds for all
Wilson loops provided their size (both width and length) is large [13, 20],
i.e. for the rectangular $R\times T$ in Fig. 1.
\begin{equation}
R, T \gg T_g
\end{equation}
where $T_g$ is the gluonic correlation length [20], defined by the range of
correlators $D(u)$ etc.

We stress at this point that the assumption (29) is very natural from the
point of view of the cluster expansion [13], and moreover it is supported by
the lattice calculations [21].

With the help of (27) one easily obtains the first term in (9); using [22]
one has
\begin{eqnarray}
E_{NP}(R) =2R \int^R_0 d\lambda \int^{\infty}_0 d\nu D(\lambda,\nu)+
\int^R_0\lambda d\lambda \int^{\infty}_0 d\nu [-2D(\lambda,\nu)+
D_1(\lambda,\nu)]
\nonumber
\\
+ \mbox{higher correlators}.
\end{eqnarray}


We note that at small $R$, $R<T_g$, [22,23]
\begin{equation}
E_{NP}(R) \sim CR^2 + 0(R^4)
\end{equation}
where $0(R^4)$ comes from higher order correlators, while at large $R$ [22]
\begin{equation}
E_{NP}(R) =\sigma R- C_0
\end{equation}
We refer the reader to ref. [24] for detailed discussion.

Now we turn to the $h$-gluon exchange diagrams of Fig. 5-10. First we must
regularize them, following the method of [25]. It was shown there  that all
divergencies of smooth Wilson loops are regularized though introduction of
some minimal distance $\delta$ on the  contour $C$ or by dimensional
regularization and renormalized) by introduction of a common  $Z$ factor
(heavy mass renormalization) and charge renormalization, generated by
divergencies in diagrams of Fig.  6,7,9,10.

We shall follow the same procedure as in [25], but we are interested not
only in the divergent parts of diagrams as in [25], but also in the large
$R$ dependence with and without background field $B_{\mu}$.

We start with the simplest $h$-gluon  exchange diagram, Fig.5, given by
(25). The first step is to use for $G_{\mu\nu}$ the Feynman-Schwinger
representation (FSR) [12]
\begin{equation}
G^{ab}_{\mu\nu}(x,y;B)= \int^{\infty}_0 ds e^{K}(Dz)_{xy}
\hat{\Phi}_{F}(x,y;B)^{ab}_{\mu\nu}
\end{equation}
where
$$K=\frac{1}{4} \int^{S}_{0}\dot{z}^2_{\mu} (\lambda)  d(\lambda);$$

\begin{equation}
\Phi_F(x,y,B) = P_FP exp ~ig \int^x_y \hat{B}_{\mu} dz_{\mu} exp~ 2~ ig
\int^S_0 \hat{F}_{\lambda\rho} (z(\lambda))d\lambda
\end{equation}
and $\Phi_F$ differs from the usual parallel transporter (18) by the ordered
(due to $P_F$) insertions of the gluon color magnetic moment interaction
$2g\hat{F}_{\lambda \rho}$. The use of FSR is the  central point of the
method since it enables one to write explicitly dependence on $B_{\mu}$ - it
enters via $\Phi_F$. As a next step  we must average (25) over the ensemble
$\{ B_{\mu}\}$, i.e. to compute an average of an expression of the form.
\begin{equation}
<W^{(2)}>_B\sim <\Phi t\hat{\Phi}_F t \Phi>_B
\end{equation}
where $\hat{\Phi}_F$ is in the adjoint while $\Phi$ in the fundamental
representation. To understand better the structure of (37) let us take the
limit $N_c \rightarrow \infty$ and represent the adjoint line in
$\hat{\Phi}_F$ by the double fundamental line, as suggested by 'tHooft
[26]. We obtain two adjacent closed contours $C_1$ and $C_2$  as shown in
Fig. 11.
When both contours large, i.e. for $R,T\gg T_g$, and $N_C \rightarrow
\infty$ one has
\begin{equation}
<W^{(2)}>_B\sim <W_{C_1}(B)>_B<W_{C_2}>_B + 0(\frac{1}{N^2_c})\sim exp
(-\sigma (S_1+S_2))
\end{equation}
where $S_1,S_2$ are minmmal areas inside contours $C_1,C_2$.

Combining now (25),(35) and (38) we obtain
\begin{equation}
-g^2<W^{(2)}>_B = -g^2\int^{\infty}_{0} ds(Dz)_{xy}dx_{\mu}dy_{\nu} exp
(-K-\sigma (S_1+S_2))\delta_{\mu\nu}
\end{equation}
(Note that we have omitted  for simplicity the $\hat{F}_{\lambda\rho}$ term
in (36), which brings about spin-orbit corrections, see Appendix 4 for
discussion.)

Eq. (39) describes propagation of a scalar particle interacting with the
contour via two strings, the world surface of which is given by $S_1, S_2$.

We now show that for our choice of contour in Fig. 5,11, with the
horizontal size $T$ much larger than vertical size $R$, the resulting
interaction tends to zero. Indeed for a given point $z(\lambda)$ on the line
of propagator, see Fig. 5, 11 the interaction does not depend on the
$z_4(\lambda), z_1(\lambda)$ which are in the plane of the contour $R\times
T$, and depends only on the transverse coordinates $z_2, z_3$. The
interaction at the "time" $\lambda$ can be written as
\begin{equation}
V\equiv\sigma(\sqrt{z_2^2 +z_3^2 +z_4^2}+\sqrt{z_2^2 +z_3^2
+(T-z_4)^2})-\sigma T
\end{equation}
where we have normalized to the case with no $h$-gluon exchange. It is easy
to see that for a generic case $T-z_4, z_4\gg z_2, z_3$ (according to (13)
we must finally take the limit $T\rightarrow\infty$) one has
\begin{equation}
V\approx\frac{\sigma}{2}(\frac{z_2^2 +z_3^2}{|z_4|}+\frac{z_2^2 +z_3^2}
{|T-z_4|})\rightarrow 0
\end{equation}

Thus we come back to the case of a free gluon exchange and obtain from (39)
\begin{equation}
-g^2\frac{<W^{(2)}>_B}{<W>_B}=-\frac{g^2 C_2(f)}{4\pi^2}\int\int\frac{dx_4
dy_4}{(x_4 -y_4)^2 +R^2}= \frac{g^2 C_2(f)}{4\pi R}
\end{equation}
In a similar way we treat the diagrams with more gluon exchanges, like in
Figs. 8, 12 and exponentiate the result (42), obtaining the lowest order
term  $(-\frac{4}{3}\frac{\bar{\alpha}_s (R)}{R})$ in (9). One should
mention that in multiple gluon exchanges like Fig. 12 there appear in general
correlated gluon pairs, triplets etc. with average distance between gluons
of the order of $(\sigma)^{-1/2}$. These correlated exchanges are not taken
into account during exponentiation and should be accounted for separately,
yielding a Yukawa-type correction to the perturbative part of $E(R)$
starting with the order $O(g^4)$. We shall discuss these corrections in a
separate publication, they do not change results of this paper concerning
large distance logarithmic behaviour.

Now we come to the fourth order diagrams, needed to renormalize $\alpha_s$
shown in Figs. 6, 7, 9, 10.

First we discuss the structure of divergencies and logarithms in case of no
background $B_{\mu}$. Referring the reader to the Appendix 2 for details, we
can e.g. write the resulting integral for the diagram of Fig. 10:
\begin{equation}
\frac{(f(R))}{R}_{Fig.10 } = \frac{-3(1-\xi)}{16\pi^2}C_2\int
\frac{d^3x}{|\vec{x}||\bar{x}-R|^3}=\frac{-3(1-\xi)}{4\pi} \frac{C_2}{R} ln
\frac{Re}{\delta}
\end{equation}
In dimentional regularization one would have
\begin{equation}
ln\frac{R}{\delta} \rightarrow \frac{C}{\varepsilon}+ ln
\mu R + const
\end{equation}
For Figs. 6 and 7 one obtains (again $B_{\mu} = 0$)
\begin{equation}
(\frac{f(R)}{R})_{Fig.6,7} = (13-3\xi) C_2 \sim
\frac{d^4r}{|\bar{R}-\bar{r}|} \prod (r^2) \rightarrow
\frac{13-3\xi}{12\pi}C_2ln\frac{R}{\delta}
\end{equation}
where $\prod(r^2)$ is the trasverse self-energy part of gluon and ghost in
the coordinate space:
\begin{equation}
\prod (r^2) =
\frac{1}{(\bar{r}^2+r^2_4)^2}
\end{equation}

Integrating over $d^4r$ we  obtain the same logarithm  as in (43).

Some more care is needed for the graph of Fig. 9; leaving details to the
Appendix 2 we mention only the final result
\begin{equation}
(\frac{f(R)}{R})_{Fig.9} = \frac{(3-\xi)C_2}{2\pi R} ln
\frac{R}{\delta}
\end{equation}
Summing (43),(45) and (47) together we obtain
\begin{equation}
f(R)= \frac{b_0}{2\pi} ln
\frac{R}{\delta}
\end{equation}
in agreement with (11) and (12).

Consider now the case of nonzero $B_{\mu}$, and let us average each of the
diagram $W^{(4)}(B)$ in Fig.6,7,9,10 over $\{B_{\mu}\}$. We again use the
limit $N_C\rightarrow \infty$ and replace each gluon line by a double
fundamental line. Choosing the gauge fixing constant $\xi=3$ we eliminate
the logarithmic part (and divergence) of  Fig. 9 and disregard it in what
follows,  since now we are only interested in the question: what happens with
logarithms in (43),(45) and (47) when the $NP$ confining field is included.

Following the same line of reasoning as we used for the diagram of Fig. 11
and imposing the confining field $B_{\mu}$ we obtain for diagrams  of Fig.
6,7 and that of Fig. 10 three surfaces in Figs. 13 and  14 respectively,
filled  with confining film due to the area law. As in (39), we must
consider again horizontal size $T$ of diagrams in  Figs.13,14 large and a
result the surfaces $S_1$ and $S_2$ do not produce any force on the gluon
line when $T\rightarrow \infty$. This is not true, however for $S_3$ , which
provides a strong confining  force between the fundamental lines which border
$S_3$.

As a  result we must consider the behaviour of the self-energy part $\prod
(x,y)$ in Fig. 13 , defined by the $FS$ representation:
\begin{equation}
\prod(x,y) \sim \int^{\infty}_0 ds \int^{\infty}_0 d \bar{s} Dz D\bar{z}
exp(-K-\bar{K} - \sigma S_3)
\end{equation}

It is easy to show ( for details see Appendix 3) that $\prod (x,y)$
behaves as $\frac{1}{(x-y)^4}$ at small $|x-y|$, $|x-y| \ll T_g$, and has an
exponential fall-off at large $|x-y|$
\begin{equation}
\prod(x,y) \sim exp (-m_1|x-y|), |x-y| \rightarrow \infty
\end{equation}

Introducing this cut-off behaviour in (46) one obtains the following
integral which replaces the original logarithmic expression in (45)
(see details in Appendix 2).
\begin{equation}
\frac{1}{R}  ln
\frac{R}{\delta} \rightarrow I(R) \equiv \int^{\infty}_{\delta} \frac{dr}{r}
e^{-m_1r} (\frac{1}{R} \theta (R-r)+\frac{1}{r}\theta(r-R))
\end{equation}

There are two regimes of behaviour of $I(R)$: the region of the asympthotic
freedom :
\begin{equation}
m_1R\ll 1 , I(R) \approx \frac{1}{R}  ln \frac{Re}{\delta}
\end{equation}
and the infrared region
\begin{equation}
m_1R\gg 1 , I(R) \approx \frac{1}{R} \int^R_{\delta}\frac{dr}{r} e^{-m_1r}
\approx \frac{1}{R}  ln \frac{1}{m_1\delta}
\end{equation}

The same situation occurs for the diagram of Fig.14. Here the cut-off
function $exp(-m|\vec{R}-\vec{x}|) $ appears under the integral in (43),
producing the  same integral $I(R)$ as in (51).

As a final result of this section we formulate the new expression for $f(R)$
taking the $NP$ confining into account (modulo constant (nonlogarithmic)
terms)
\begin{equation}
\frac{1}{R} f_{NP}(R) = \frac{b_0}{2\pi} I(R)
\end{equation}
with $I(R)$ defined in (51).

\section{Renormalization of $\alpha_s$ in the definition ii)}

 In the second nonperturbative definition of $\alpha_s$ through the
 two-point function $H(x,y)$ (19), depicted in Fig. 4, we can again exploit
 the $N_c \rightarrow \infty$ limit. In this case each adjoint line,
 including $\hat{\Phi}(x,y,B)$, is represented as double fundamental line
 [26] and each closed fundamental contour $W_c(B)$, after averaging over
 fields $\{B_{\mu}\}$, acquires the area-law factor $exp(-\sigma
 S_{min}(C))$ - in agreement with our assumption (29). Thus each of the
 diagrams of Fig. 4  can be depicted as in Fig. 15, the difference between
 (a) and (b) is that for the gluon Green's function there are in addition
 gluon spin operators inserted on the gluon line, leading to the
 spin-dependent effects, which we disregard in the first approximation.

 The diagrams of Fig. 4  can be written in the form similar to (49),
 which is  discussed in  detail in the Appendix 3
 \begin{equation}
 H_{\mu\nu} (x,y) = \frac{g^2}{(4\pi)^2} b_0
 (\partial_{\mu}\partial_{\nu}- \partial^2 \delta_{\mu\nu})
 \bar{\prod}(x,y)
 \end{equation}
 where $\bar{\prod} $ is defined as
 \begin{equation}
 \bar{\prod}(x,y)= \int^{\infty}_0 ds\int^{\infty}_0d\bar{s}~e^{-K-\bar{K}}
 (Dz)_{xy}(D\bar{z})_{xy} ~e^{-\bar{V}}
 \end{equation}
 with
 \begin{equation}
\bar{V} = S^{(1)}_{min}+S^{(2)}_{min}+S^{(3)}_{min}
 \end{equation}
 and $S^{(i)}_{min}$ correspond to the minimal areas inside the 3
 contours depicted in Fig.15.

 We have omitted in (57) for simplicity the gluon spin terms in the
 Green's function $G$ in (19).

 At small distances $|x-y|$ one can follow analysis of Appendix 3 and
 obtain that
 \begin{equation}
 \bar{\prod}(x,y) \sim \frac{1}{(x-y)^{2d-4}}
 \end{equation}

 This leads to the familiar answer for the Fourier-transform of $H_{\mu\nu}$
 [11]
\begin{equation}
 H_{\mu\nu}(p) = \frac{g^2 b_0}{(4\pi)^2}(p_{\mu} p_{\nu} - p^2
\delta_{\mu\nu})(N_{\varepsilon}- \ln \frac{p^2}{\mu^2} + const)
\end{equation}
with
$$b_0 = \frac{11}{3}N_c\;,\; N_{\varepsilon }= \frac{1}{\varepsilon} - C +
\ln 4\pi\;, $$
$C$ is the Euler constant,  $C=0.577$.

At large $\mid x-y \mid$ one must take into account the confining dynamics
present in $\bar{V}$ and find the corresponding mass. To this end one must
look at the expressions (55), (56) more carefully.

To understand the structure of $H(x,y)$, one can dissect the graph in Fig.
15 along the dash-dotted line  and one obtains a system of two gluons
1 and 2 connected by a pair of strings, as shown in Fig. 16. One of the
strings goes through the fixed point 3 which appears due to the dissection
of the straight-line parallel transporter $\Phi(x,y;B)$ -- see Fig. 15.

One can show that the mass of the system of two gluons with two strings in
the configuration of Fig. 16 is larger than the corresponding mass without
the additional condition of  string passing through the point 3:
\begin{equation}
m(1,2;3) > m(1,2) \equiv m_2
\end{equation}

Thus we come to the conclusion that $H$ describes the Green's  function of a
two-gluon glueball with a quantum nember $J^{PC} = 1^{-+}$ (and with an
additional "heavy gluon" propagating along the straight line of
$\hat{\Phi}(x,y,B)$, if we keep the point 3 in Fig. 16, and without it if we
calculate a more light system according to the inequality (60)).

An estimate of $m(1,2)$ with $J^{PC} = 1^{-+}$ has been done within the
proper time Hamiltonian in [27]
\begin{equation}
m(1,2) \cong 2 GeV \pm 0.15 GeV
\end{equation}
A similar estimate has been obtained in the lattice calculations, where
usually two neighbouring members of the same multiplet $L=1, S=1$ are
studied [28].

Using the arguments given in the Appendix 3, one can approximate
$\bar{\prod}(x,y)$ as
\begin{equation}
\bar{\prod}(x,y) \simeq C_0\frac{exp(-m_2 \mid x-y
\mid)}{(x-y)^{2D-4}}
\end{equation}
which after renormalization yields  the Fourier-transform
\begin{equation}
\bar{\prod}(p) \approx C_0 \ln\frac{m^2_2+p^2}{\mu^2}+const
\end{equation}

A comparison of (59) and (63) shows that up to power corrections two
expressions  have the same $\mu$ dependence, and hence the same $Z$
factors. Actually all the change from (59) to (63) can be described as
introduction of a new (fifth) external momentum component $p^2_5 \equiv
m^2_2$, so that the total momentum squared is in case of nonperturbative
background equal to $p^2+m^2_2$.
This fact does not change the Gell-Mann-Low equations which to one-loop
order are [15]
\begin{equation}
\frac{dg}{d\ln\mu} =-b_0 \frac{g^3}{16\pi^3}
\end{equation}
and solving (64) with boundary conditions as in (55) and (63) (neglecting
power terms as before for large $(m^2+p^2) x^2_0\gg 1)$ one obtains
\begin{equation}
\alpha_s (p) =\frac{\alpha_s(\mu)}{1+\frac{b_0}{4\pi}\alpha_s(\mu) \ln
\frac{m^2_2+p^2}{\mu^2}}
\end{equation}

Introducing the $\Lambda$ parameter (65) is rewritten as
\begin{equation}
\alpha_s (p) =\frac{4\pi}{b_0 \ln\frac{m^2_2+p^2}{\Lambda^2_{NP}}}
\end{equation}

We have used notation $\Lambda_{NP}$ to stress that this new $\Lambda$
parameter may differ from the free vacuum $\Lambda_{QCD}$ and needs a new
determination from the experimental data. We stress here again that the
appearence of $m^2_2$ together with $p^2$ is possible because of
renormalization invariance of $gB_{\mu}$ [11],which in its turn yields
invariance of $\sigma$ and $m_2\sim =\sqrt{\sigma}$.

At large $p$ one  obtains in (66) the familiar $AF$ logarith. This logarithm
is still large for $p=0$ if one chooses a low normalization parameter
$\Lambda_{NP}$,$ \Lambda_{NP} \sim \Lambda_{QCD}$. Thus the $AF$ phenomenon
can be maintained down to very low $p$ in the Euclidean region. This is
similar to the situation which we have observed in the Wilson-loop
renormalization scheme, eg. (53).

There, however, the mass $m_1$ corresponds to the one-Wilson-loop dynamics,
while in our case actually a doubled film defines dynamics for $m(1,2)$,
with $\sigma_a \cong 2\sigma$, so that the mass $m_2 \cong \sqrt{2} m_1$.
This fact demonstrates that the $IR$ cutoff is scheme dependent. (Note that
our definition of the renormalization scheme differs from the usual one,
and the scheme dependence actually means the dependence on the physical
process in question).

\section{Renormalons and asymptotics of the perturbative series}
In this section we study the consequences of the modified $AF$ (66) for the
properties of  the PTH series . We shall concentrate for definiteness on the
Euclidean correlator of e.m. currents $\prod^{e.m.}(Q^2)$ as
in [29, 30] and shall write it as in [30] in a form
exhibiting the contribution of the renormalon [5]
\begin{equation}
\prod^{e.m.}(Q^2)
=\frac{1}{2\pi^2b_0}(\frac{1}{\alpha_s(\mu^2)}-\frac{1}{\alpha_s(Q^2)} )+
\frac{1}{2\pi^2b_0} \ln (\frac{\alpha_s(Q^2)}{\alpha_s(\mu^2)})+\sum
\tilde{p}_n\alpha_s^n(Q^2) + \Delta\prod
\end{equation}
where we have separated out of the PTH series the contribution of the set of
diagrams shown in Fig. 17, denoted by $\Delta \prod$.
 We shall first discuss the infrared ($IR$) renormalon, [5], i.e. the
 contribution from the graphs of Fig.17 from the domain of integration over
 $k$ with $k\ll Q$. If the coupling constant $\alpha_s$ is normalized at
 $Q^2$ (we shall take it at $Q^2+m^2$ and (66) is used for $\alpha_s$, we
 have
\begin{equation}
\Delta \prod=
\frac{\alpha_s(Q^2)}{8\pi^3}\sum_n(\frac{b_0\alpha_s(Q^2)}{4\pi})^n
\int^{Q^2}_0 \frac{k^2dk^2}{Q^4}\ln^n (\frac{Q^2+m^2}{k^2+m^2})
\end{equation}

The analysis of the integral (68) made in the Appendix 5, yields

\begin{equation}
I=
\int^{Q^2}_0
\frac{k^2dk^2}{Q^4}\ln^n (\frac{Q^2+m^2}{k^2+m^2}) =\left\{
\begin{array}{ll}
\frac{n!}{2^{n+1}}~~, & n<2n_0\\
\eta\frac{n_0^{n+1}}{n+1}~~, &n>2n_0,~~\eta<1
\end{array}
\right.
\end{equation}
with the notation $n_0=\ln \frac{Q^2+m^2}{m^2}$.

As a result we obtain the following contribution of the graphs in Fig.17
\begin{equation}
\Delta \prod = \frac{4\pi}{8b_0\pi^3}\sum^{\infty}_{n\gg 1}
(\frac{\alpha_s(Q^2)b_0}{8\pi})^n q_n
\end{equation}
where
\begin{equation}
q_n=\left \{
\begin{array}{ll}
(n-1)!~~, & n<2n_0\\
\eta \frac{(2n_0)^{n}}{n}~~, &n>2n_0
\end{array}
\right.
\end{equation}

We remind that the estimate (71) is an upper limit for $q_n$.

First we discuss a connection with the previous results [29,30] on the IR
renormalons. In this case $m^2\rightarrow 0$ (no background field) and
$n_0\rightarrow \infty; q_n$ is defined as in (71) for any $n$ and the Borel
transform of $\Delta \prod $ is as in [29,30]
\begin{equation}
\Delta \prod = (16\pi^3)^{-1} \int^{\infty}_0
\frac{exp(-\frac{t}{\alpha_s(Q^2)})dt}{1-\frac{tb_0}{8\pi}}
\end{equation}
the Borel integral (72) exibits IR renormalon pole at $t=\frac{8\pi}{b_0}$
which makes the Borel summation of the PTH series prohibitive.


In our case, when $n_0$ is finite, the situation is drastically changed.
Namely, having in mind that the estimate (71) is an upper limit of $q_n$
both for $n>2n_0$ and for $n<2n_0$, we can take the sum in (70) which does
not diverge and even does not need Borelization. Leaving details for the
Appendix 5 we quote only the final result
\begin{equation}
\Delta
\prod = \frac{\alpha_s}{8\pi^3} \sum^{\infty}_{n=1}
\int^{n_0}_0\frac{\alpha_s(Q^2)tb_0}{4\pi})^n e^{-2t} dt \approx
\frac{b_0\alpha_s^2(Q^2)}{32\pi^4}
\end{equation}
There are no poles in the plane of the Borel variable, as can
be easily seen from the absense of the factorial growth of the coefficient
$q_n$ at arbitrarily large $n$. We thus conclude that the convergence of PTH
series is radically improved when the strong nonperturbative background is
taken into account, and in particular the IR renormalons disappear.

We now turn to the ultraviolet (UV) renormalons [5]. Those come from the
same  series of graphs, Fig.17, but from the integration region with $k^2\gg
Q^2$. It is easy to understand that the UV renormalons do not change due to
the modified AF logarithm (66), since in this kinematical region $k^2\gg
m^2$ and the change is immaterial. Hence we have as in [30] the estimate
\begin{equation}
\Delta \prod(UV ) = -\frac{1}{6\pi^3}
\sum_{n}
\alpha_s^n (-\frac{b_0}{4\pi})^{n-1} (n-1)!
\end{equation}
which brings a pole at negative value of the Borel parameter,
$t=-\frac{4\pi}{b_0}$, and hence the Borel sum of the PTH series is well
defined.

\section{Conclusion}

We have introduced in this paper a new perturbation theory, based on the
explicit separation of $NP$ and perturbative fields, Eq.(1).

The former are defined by their vacuum correlators as in (27), which at
large distances enter only as one constant - the string tension $\sigma$
(30).

This is the input of the theory. The PTH expansion in powers of
perturbative field $a_{\mu}$ is then defined unambigiously in terms of the
$NP$ input and coupling constant $\alpha_s$.

The main emphasis of the paper is the drastic change in the behaviour of
renormalized $\alpha_s$ at large distances as compared to the usual (free)
PTH.

Two different $NP$ definitions of the running coupling constant are used,
Eqs.(9,13) and (19), and in both cases the one-loop contribution is shown to
be bounded at large distances , in contrast to the logarithmic growth of the
free one-loop result.

It is important  that  Gell-Mann-Low equations are not changed and the
resulting behaviour of $\alpha_s(p)$ is characterized by only one new
parameter, $m^2$ as in (65-66), which is RG  invariant. The  mass $m$ depends
on the process  and in the typical situation of a gluon exchange in the
$e^+e^-$ annihilation, is of the order of one GeV. As a result the magnitude
of $\alpha_s(p)$ everywhere in the Euclidean region of $p^2$ is limited and
the Landau ghost pole is not inevitable. Rough estimates show that
$\alpha_s(p)$ is less than unity in the Euclidean region; enabling one to
use perturbation expansion at all distances.

As a consequence there appears a solution of the long standing problem of IR
renormalons. It is shown in Section 6, that IR renormalons are absent when
the running coupling has the form (66). Hence the Borel summation of the new
perturbative series is not prohibited by the IR renormalons, and one
encounters in the right Borel-parameter half-plane only the distant
singularities  due to instanton-antiinstanton $I\bar{I}$ pairs.

We do not consider here important questions, e.g. how inclusion of
$I\bar{I}$ pairs into the $NP$ vacuum fields influences the properties of
the new perturbative series.

We also do not discuss the behaviour of $\alpha_s(p)$ in the Minkowskian
region, $ p^2<0$,which shall be presented in a subsequent paper.

During the earliest stage of this study the author intensely discussed with
P. van Baal and G.'tHooft. Their useful remarks and suggestions are
gratefully acknowledged.

The author is grateful for useful discussions to A.M.Badalyan, K.
G.Boreskov, A.Yu.Dubin, B.L.Ioffe, A.B.Kaidalov, L.N.Lipatov, M.I.Polikarpov
and K.A.Ter-Martirosyan.

\newpage

\underline{\bf APPENDIX  1}\\

{\bf The contribution of the term $L_1$}\\

\setcounter{equation}{0}
\def\theequation{A1.\arabic{equation}}

Let us calculate the contribution of the term $L_1$ to the gluon propagator.
If one denotes by $<>_a$ the integral $Da_{\mu}$ with the weight $L(a)$ as
in (2), we obtain
\begin{eqnarray}
\Delta^B_{\mu_1 \mu_2}(x_1,x_2) \equiv <a_{\mu_1}(x_1) a_{\mu_2}(x_2)>_a=
\nonumber \\
\int d^4y_1 d^4y_2 G_{\mu_1\nu_1}(x_1,y_1)D_{\rho}F_{\rho \nu_1}(y_1)
D_{\lambda} F_{\lambda\nu_2}(y_2) G_{\nu_2\mu_2} (y_2,x_2)
\end{eqnarray}
The gluon Green's function $G_{\mu\nu}$ is given in (8) and depends on the
background field $B_{\mu}$, as well as $D_{\mu}$ and $F_{\rho\lambda}$.
Averaging over $\{ B_{\mu}\}$ is done in (A.1.1) using cluster expansion .
To get a simple estimate of $\Delta$ we replace $G_{\mu\nu}$ by the free
Green's function $G^{(0)}_{\mu\nu}$ and take into account that
\begin{eqnarray}
<D_{\rho}F_{\rho \nu_1}(y_1)
D_{\lambda} F_{\lambda\nu_2}(y_2)>_B\Rightarrow \frac{\partial}{\partial
y_{1\rho}}\frac{\partial}{\partial y_{2\lambda}}
<F_{\rho\nu_1}(y_1)F_{\lambda\nu_2}(y_2)>
\end{eqnarray}
and for the latter we use the representation [13]  in terms of two
independent Lorentz structures, $D(y_1-y_2)$ and $D_1(y_1-y_2)$. The
contribution of $D$ is (that of $D_1$ is of similar character) in the
momentum space
\begin{eqnarray}
\Delta^B_{\mu_1\mu_2}(k) \sim
\frac{(k^2\delta_{\mu_1\mu_2}-k_{\mu_1}k_{\mu_2})}{k^4}D(k)
\end{eqnarray}
where
\begin{eqnarray}
D(k) = \int d^4y e^{iky}D(y)
\end{eqnarray}
Insertion in (A.1.4)of the exponential fall-off for $D(y)$ found in lattice
calculations [31] yields
\begin{eqnarray}
D(k) = \frac{<F^2(0)>}{(N^2_c-1)}\cdot
\frac{\pi^2\mu}{(\mu^2+k^2)^{5/2}}~.
\end{eqnarray}

With $\mu\approx 1 GeV,~~D(k)\approx 0.12$. Thus $\Delta^B(k)$ is a soft
correction fast decreasing with $k$ as $k^{-5}$ to the perturbative gluon
propagator.

\newpage

\underline{\bf APPENDIX 2}\\

{\bf Calculation of the Wilson-loop charge renormalization}\\

\setcounter{equation}{0}
\def\theequation{A2.\arabic{equation}}

Here we discuss in more detail calculation of graphs, corresponding to
Figs.6-10, which contribute to the charge renormalization in the "Wilson"
renormalization scheme, Eq.(9).

We start with the graph of Fig.6. A typical term contributing to the
amplitude can be written as
\begin{eqnarray}
J^{(6)}& =& \int \Delta_{\mu_1 \mu}(y_1,x) \frac{\partial}{\partial
x_{\nu}}
\Delta_{\mu\lambda}(x-y)\Delta_{\nu\rho}(x-y)
\frac{\vec{\partial}}{\partial y_{\rho} } \Delta_{\lambda\mu_2}(y,y_2)\times
\nonumber
\\
&\times & dy_{1\mu_1}dy_{2\mu_2} d^4x d^4y
\end{eqnarray}

Let the upper and lower horizontal lines in Fig.6 correspond to $y_{14}\in
[0,T], y_{1i}=0$ and $y_{24}\in [0,T], y_{21}=R$, respectively. According to
our discussion in Section 4,  the first and the last gluon propagators in
(A 2.1) are not affected by confinement when $T\rightarrow \infty$ and one
obtains in (A 2.1)
\begin{eqnarray}
J^{(6)}\sim dx_4 \frac{d^3x}{|\vec{x}|} \partial_{\nu}
\prod(x,y)\vec{\partial}_{\nu} \frac{d^4y}{|\vec{R}-\vec{y}|}.
\end{eqnarray}
Applying Fourier transformation to both propagators in (A 2.2) yields
\begin{eqnarray}
J^{(6)}\sim T\frac{d^4r\prod (r)}
{|\vec{R}-\vec{r}|}
\end{eqnarray}

In case of no background field $\prod(r)$ is simply a product of two free
gluon propagators and one gets
\begin{eqnarray}
J^{(6)}\sim \frac{Td^3\vec{r}}
{|\vec{R}-\vec{r}||\vec{r}|^3}\sim \frac{T}{R}\ln \frac{R}{\delta}
\end{eqnarray}
where $\delta^{-1} $ is the $UV$ cut-off parameter.

We introduce the following notation for the one-loop renormalization of
$\alpha_s$ in the scheme (i)
\begin{eqnarray}
\alpha_s(R) = \alpha_s^{(0)} (1+f(R)\alpha_s^{(0)}+...)
\end{eqnarray}
In the case of no background the diverging contribution of graphs Fig.6 and
7 to $f(R)$ is [25]
\begin{eqnarray}
\stackrel{(6,7)}{f(R)} =
\frac{13-3\xi}{12\pi}C_2\ln\frac{R}{\delta}
\end{eqnarray}

This should be compared to the momentum space calculation [17] of no
background case (Feynman gauge, $\xi = 1$)
\begin{eqnarray}
\alpha_s(Q) = \alpha^{(0)}_s(1+\frac{\alpha^{(0)}_s}{4\pi}[(-\frac{5}{3}
C_2)(-\frac{2}{\varepsilon}+C - \ln 4\pi + \ln\frac{Q^2}{\mu^2}) +
\frac{31}{9} C_2]
\end{eqnarray}
where $C= 0.577$ in the Euler constant and in the $\overline{MS}$
scheme one should omit the factor $(\frac{2}{\varepsilon} + C- \ln
4\pi)$ in (A.2.7). We now consider the case of nonzero background field, and
take $\prod(r)$ as estimated in Appendix 3, Eq. (A3.21)
\begin{eqnarray}
\prod(r) \approx \frac{exp(-m_1\sqrt{\bar{r}^2 + r^2_4})}{(\bar{r}^2 +
r^2_4)^2}
\end{eqnarray}
Insertion of $\prod(r)$ (A2.8) into (A2.3) yields instead of (A2.4)
\begin{eqnarray}
J^{(6)} = \frac{1}{4\pi}\int
 \frac{Td^3\vec r}{\mid \bar{R} - \bar{r} \mid \mid \bar{r}
\mid ^3} \int^{\infty}_{0} \frac{e^{-m_1r chx}dx}{(ch x)^3} = TI(R)
\end{eqnarray}
where
\begin{eqnarray}
I(R) = \int^{\infty}_0
\frac{dx}{(ch x)^3}\int^{\infty}_{\delta}\frac{dr}{r} e^{-m_1r ch
x}(\frac{1}{R} \Theta(R-r)+\frac{1}{r} \Theta(r-R))
\end{eqnarray}
It is clear that the integral over $d x$  converses at $x \sim 1$ and
therefore one can replace in the integral over $r, \;\;m_1 ch x \rightarrow
m^*, m^* > m_1$. It is interesting to study two limiting cases:
 1) $m^* R\ll
1$~~~~and  2)$ m^*R \gg 1$.
One has
\begin{eqnarray}
I(R) \approx \frac{1}{R} \ln \frac{Re}{\delta}\;, \; m^*R \ll 1
\nonumber
\\
I(R) \approx \frac{1}{R} \ln \frac{1}{m^*\delta}\;, \; m^*R \gg 1
\end{eqnarray}
We turn now to the grphs of Fig. 8 and 9. Due to the nonabelian character of
fields the graph of Fig. 9 is proportional to
\begin{eqnarray}
tr(t^a t^b t^a t^b) = (\frac{N^2 - 1}{2N})^2 N-\frac{N}{4}(N^2-1)
\end{eqnarray}
The first term on the r.h.s. of (A2.12) can be considered as the "abelian
extrapolation" of the graph of Fig. 8 to the  crossed kinematics of Fig. 9,
while the second term on the r.h.s. of (A2.12) represents the pure
nonabelian contribution. In the case  of no background one obtains
therefore the following nonabelian contribution to the Wilson loop from the
graph of Fig. 9
\begin{eqnarray}
\frac{1}{N} W^{(2)} = -\frac{g^4(N^2-1)}{(4\pi^2)^28} \int \frac{dy_{1\mu}
dy_{2\mu} dx_{1\nu} dx_{2\nu}}{(y_1-y_2)^2(x_1-x_2)^2}
\end{eqnarray}
which gives the following divergent contribution to $f(R)$ (cf(A2.5))
\begin{equation}
f^{(9)}(R) = \frac{(3-\xi)C_2}{2\pi R}\ln\frac{R}{\delta}
\end{equation}
This should be compared with the momentum space calculation of [17] (we
keep only divergent (logarithmic) term)
\begin{equation}
\Delta \alpha_s(Q) = -\frac{(\alpha_s^{(0)})^2}{4\pi}\frac{N^2-1}{2N}N
\cdot(3-\xi) \ln \frac{Q^2}{\lambda^2}
\end{equation}

To simplify our discussion we shall use now the gauge $\xi = 3$ and
therefore one can disregard the graph of Fig. 9.
This is true also in case of nonzero background, since we are interested
here only in the modification of the divergent term $\ln \frac{R}{\delta}$
due to the background.

We now turn to the graph of Fig. 10.

Its contribution to $f(R)$ is
\begin{equation}
\frac{f^{(10)}(R)}{R} = -\frac{3(1-\xi)}{16\pi^2} C_2 \int \frac{d^3x}{\mid
\vec{x} \mid \mid \vec{x}- \vec{R} \mid^3}
\end{equation}
where $\mid \vec{x} \mid^{-1}$ enters from the static gluon propagator
$\Delta(y_1-x)$ while the favctor $\mid \vec{x} - \vec{R} \mid^{-3}$ is from
the triangle part $\Gamma(x,y_2, y_3)$ of Fig. 10. In case of the confining
background one has to insert in the integral in (A2.16) the factor,
$exp(-m_1 \mid \bar{x} - \bar{R} |)$, which yields again the function $I(R)$
as in (A2.9).  This factor appears in the asymptotics of $\Gamma(x,y_2,
y_3)$ due to the confining area law with the surface $S_3$, shown in Fig.
14. The asymptotics holds true when $\mid \bar{x} - \bar{R} \mid$ is large
and the mass $m_1$ is the same as in (A2.8), since at  large $\mid x - y_2
\mid\;,\; \mid x - y_3 \mid$ and fixed $\mid y_2 - y_3 \mid$ the function
$\Gamma(x,y_2, y_3)$ behaves as the two-point function $\prod(x,y)$
discussed in Appendix 3.

Combining all contributions from all graphs of Fig. 6-10 one obtains finally
\begin{equation}
\frac{1}{R}f^{(total)}(R) = \frac{b_0}{2\pi} I(R)\;, \; b_0 = \frac{11}{3}N
\end{equation}

For small $R\;, m_1R \ll 1$, one has due to (A2.11) the usual
renormalization factor
\begin{equation}
\alpha_s(R) = \alpha_s^{(0)}(1+\alpha_s^{(0)} \frac{b_0}{2\pi} \ln
\frac{R}{\delta}+ ...)
\end{equation}
while for large $R$, $m_1R \gg 1,$ the function $RI(R)$ tends to a constant
and the unlimited growth of the renormalized charge $\alpha_s(R)$ is frozen.

\newpage

\underline{\bf APPENDIX 3}\\

{\bf Properties of the self-energy function $\prod(x,y)$}\\

\setcounter{equation}{0}
\def\theequation{A3.\arabic{equation}}

In this appendix we study the properties of the self-energy part
$\prod(x,y)$, Fig.~~~, both in the coordinate and in the momentum space. We
consider for simplicity spinless particles in the loop, which is true for
the ghost loop, inclusion of spin is discussed in  Appendix 4.

The $FSR$ for $\prod(x,y)$ is [12]
\begin{equation}
\prod(x,y) = \int^{\infty}_0 ds \int^{\infty}_0 d\bar{s} (Dz)_{xy}
(D\bar{z})_{xy} e^{-K-\bar{K}-V}
\end{equation}
where $V$ reduces to $\sigma S_{min}$ for large loops [13]
\begin{equation}
 K=  \int^{s}_0
 \frac{\dot{z}^2_{\mu}(\lambda) d\lambda}{4}~~,~~~ \bar{K}=
\int^{\bar{s}}_0\frac{\dot{\bar{z}}^2_{\mu}(\bar{\lambda})d\bar{\lambda}}{4}~~;
\end{equation}

Introducing as in [24] "center-of-mass" and relative coordimates, $R_{\mu}$
and $r_{\mu}$
\begin{equation}
R_{\mu} =  \frac{\bar{s}}{s+\bar{s}}z_{\mu}(\lambda) +
\frac{s}{s+\bar{s}}\bar{z}_{\mu}(\lambda)~~; ~~~~
r_{\mu}= z_{\mu}(\lambda)-\bar{z}_{\mu}(\lambda)
\end{equation}
and new variables $\mu, \bar{\mu}$ instead of $s,\bar{s}$
\begin{equation}
\mu =  \frac{T}{2s}~~, ~~~~\bar{\mu}= \frac{T}{2\bar{s}}~~,
{}~~~\tilde{\mu} = \frac{\mu\bar{\mu}}{\mu+\bar{\mu}}
\end{equation}
we obtain
\begin{equation}
\prod(x,y) = T^2 \int^{\infty}_0\frac{d\mu}{2\mu^2} \int^{\infty}_0
\frac{d\bar{\mu}}{2\bar{\mu}^2} DR\cdot Dr e^{\{-\int^T_0 d\tau \frac{(
\mu+\bar\mu)}{2}\dot{R}^2 + \frac{\tilde{\mu}}{2} \int^T_0 \dot{r}^2 d\tau
+V}\}
\end{equation}
where $T\equiv |x-y|$.
The boundary conditions for trajectories $R_{\mu}(\tau),r_{\mu}(\tau)$ are:
\begin{equation}
R_{\mu}(0) = y_{\mu}; ~~R_{\mu}(T) = x_{\mu}; r_{\mu}(0) = r_{\mu}(T)=0
\end{equation}
as disscussed in [24], one can replace $V=\sigma S_{min}$ by the expression
$V=\sigma|\vec{r}(\tau)|$, with the accuracy of $\sim 5\%$ for the mass
eigenvalues. In this case one can go over to the Hamiltonian form in (A
3.5) using the relation
\begin{equation}
\int DR~Dr~e^{-\int L d\tau}= <R=x, r=0| e^{-\int H d\tau}|R=y; r=0>
\end{equation}
Since the motion in $R_{\mu}$ and $r_4$ coordinates in free one can easily
calculate its contribution and the result for $\prod(x,y)$ is
\begin{equation}
\prod (x,y) = \frac{1}{4(2\pi)^{5/2}} \int\int^{\infty}_0 \frac{d\mu_1 d\mu_2
e^{-\frac{\mu_1+\mu_2}{2}T}}{\tilde{\mu}^{3/2}\sqrt{T}} G(0,0;T)
\end{equation}
where $G(r_f,r_{in};T)$ is the Green's function of relative quark-antiquark
(or gluon-gluon) motion,
\begin{equation}
G(0,0;T) = \sum_n C_n(\tilde{\mu}) e^{-\varepsilon_n(\tilde{\mu})T} ,
\end{equation}
$C_n(\tilde{\mu}) = |\psi_n(0)|^2, \psi_n$ and $\varepsilon_n(\tilde{\mu})$
are  eigenvalues of the Hamiltonian
\begin{equation}
H\psi_n = \varepsilon_n\psi_n; H= \frac{\vec{p}^2}{2\tilde{\mu}}+\sigma
|\vec{r}|~.
\end{equation}
For linear potential $C_n(\tilde{\mu})=\frac{\tilde{\mu}\sigma}{2\pi}$ and
does not depend on $n$ , while $\varepsilon_n(\tilde{\mu})$ is
\begin{equation}
\varepsilon_n(\tilde{\mu}) = (2\tilde{\mu})^{-1/3} \sigma^{2/3} a(n)
\end{equation}
where $a(n)$ are numbers, obtained from the solution of the dimensioless
equation, with the asymptotics
\begin{equation}
a(n) = (\frac{3\pi}{2})^{2/3}(n+\frac{1}{2})^{2/3}~~, ~~~n\rightarrow \infty
\end{equation}
One may get a false impression that expressions (A 3.9- A 3.12) are
nonrelativistic. In fact relativism is contained here in the
$\tilde{\mu}$ variable which finally yields
spectrum very close to that of the
relativistic quark model [24] or $1+1$ QCD.

Insertion of (A 3.9) into (A 3.8) yields
\begin{equation}
\prod(x,y) = A\sum_n \int\frac{d\mu_1 d\mu_2}{\sqrt{\tilde{\mu}T}}e^{-M_n
(\mu_1, \mu_2)T}
\end{equation}
where
\begin{equation}
M_n (\mu_1, \mu_2)= \frac{\mu_1+\mu_2}{2} + \varepsilon_n(\tilde{\mu})~,
\end{equation}
with $$A\equiv \frac{\sigma}{4(2\pi)^{7/2}}$$

At large $T\equiv |x-y|$ the asymptotics of (A 3.13) is obtained performing
the $d\mu_1, d\mu_2$ integrations via the steepest descent method  [24],
which yields
\begin{equation}
\prod(x,y) = \frac{A\pi\sqrt{3}}{T^{3/2}} \sum_n\sqrt{\bar{M_n}}
e^{-\bar{M}_nT}
\end{equation}
where $\bar{M}_n = 4\sqrt{\sigma} (\frac{a(n)}{3})^{3/4}$ is obtained from
the minimum of $M_n(\mu_1,\mu_2)$ in variables $\mu_1, \mu_2$.

Therefore at large $T$ the correlator $\prod$ decays exponentially
\begin{equation}
\prod(x,y)\sim  \frac{exp(-\bar{M}_0T)}{T^{3/2}}
\end{equation}
where $\bar{M}_0$ is the minimal mass value.

At small $T$ one should  use (A 3.13) and take into account , that in this
situation the sum can be replaced by the integral
\begin{eqnarray}
\sum_n e^{-\varepsilon_n(\tilde{\mu})T}\rightarrow \int dn
e^{-\lambda(\tilde{\mu})T(n+\frac{1}{2})^{2/3}}=\nonumber \\
=\frac{3}{2} \frac{\Gamma(\frac{3}{2})}{(\lambda T)^{3/2}}~~,~~~~
\lambda =(2\tilde{\mu})^{-1/3} \sigma^{2/3}(\frac{3\pi}{2})^{2/3}~~.
\end{eqnarray}
            Insertion of (A 3.17) into (A 3.13) yields
\begin{eqnarray}
\prod (x,y) =
\frac{1}{4(2\pi)^4T^2}\int^{\infty}_0d\mu_1\int^{\infty}_0 d\mu_2
e^{-\frac{(\mu_1+\mu_2)T}{2}} = \frac{1}{16\pi^4T^4}
\end{eqnarray}

This is the exact answer for the free scalar gluon loop diagram. It is
remarkable that the sum over bound states reproduces the correct small
distance behaviour.

Finally we shall write the answer  for the momentum space. Choosing
$p=(\vec{0},p_4)$ one has
\begin{eqnarray}
\prod (p) = \bar{A} \sum_n \int^{\pi/2}_{-\pi/2} \cos^2\theta d\theta \sum_n
\frac{\sqrt{M_n}}{(M_n-
ip_4\sin
\theta)^{D-3/2}}
\end{eqnarray}
with $M_n= m\sqrt{n+\frac{1}{2}},~~~ m= \frac{4\sqrt{\sigma}}{3^{3/4}}
(\frac{3\pi}{2})^{1/2}, ~~ \bar{A}= \frac{\sqrt{3}\sigma}{(2\pi)^{3/2}}
\Gamma(D-\frac{3}{2})$.

At large $n$ one can replace the sum by the integral of the type
$(D=4+\varepsilon)$.
\begin{eqnarray}
\int^{\infty}_{m_0}\frac{dM}{(M-ia)^{D-3}}= \frac{1}{\varepsilon} - \ln
(m_0-ip_4\sin\theta)
\end{eqnarray}

Integration over $d\theta$ of the logarithmic term in (A 3.20) yields
finally the effective logarithm which appears in (63)
\begin{eqnarray}
\int^{\pi/2}_{-\pi/2}\cos^2 \theta d\theta \cdot
\ln (m_0-ip_4\sin\theta) = \frac{\pi}{4} \ln (m^2_0+p^2_4
\sin^2\bar{\theta})
\end{eqnarray}

On another hand each term with a given $n$ contributes a simple pole in
$\prod(p)$, which can be separated out looking more closely at the
integration region near $\theta \approx \pm \frac{\pi}{2}$. Indeed writing
$\sin \theta = \cos \eta \approx 1 -\frac{\eta^2}{2}$ one can rewrite the
integral in (A 3.19) as
\begin{eqnarray}
\int^1_0\frac{\eta^2 d\eta}{(\delta+ M_n\frac{\eta^2}{2})^{5/2}} \approx
\frac{1}{\delta} (\frac{2}{M_n})^{3/2}\cdot\frac{1}{3}
\end{eqnarray}
where $\delta = M_n - ip_4$. Finally the contribution of the poles is
\begin{eqnarray}
\prod (p) = \sum_n\frac{\sqrt{3}\sigma}{2\pi (M^2_n+p^2)}
\end{eqnarray}

The sum (A 3.23) is diverging for large $n$, calling for the regularization
which we have explicitly done in (A 3.20).

The fact that $\prod(p)$ has only simple poles equidistantly placed at
$-p^2=M^2_n=m^2(n+\frac{1}{2})$ tells us that the difference $\prod(p)
-\prod(p_0)$ is proportional to the $\psi$ function, $\psi(z) =
\frac{\Gamma'(z)}{\Gamma(z)}$.

Indeed, one can write
\begin{equation}
\prod(p) - \prod(p_0)\sim \psi (\frac{p^2+\frac{1}{2}m^2}{m^2})
\end{equation}
which asymptotically at large $\frac{p^2}{m^2}\gg 1 $ behaves as
\begin{equation}
\psi(z) \sim \ln z - \frac{1}{2z} - \sum^{\infty}_{n=1} \frac
{B_{2n}}{2nz^{2n}}
\end{equation}
where $B_n$ are Bernulli numbers.
In this way we again obtain the logarithmic term similar to (A 3.21).

Finaly we derive one more form of $\prod(x,y)$ which also was used in the
main text. One can rewrite (A 3.15) separating in $\bar{M}_n$ the
constant term, not depending on $n$:
\begin{equation}
\bar{M}_n= m_0+M(n)
\end{equation}
such that at $n= n_{min}, M(n_{min})=0$.

 Having in mind that $d\bar{M}^2_n = m^2 dn$, one can rewrite (A 3.15)
 as
 \begin{equation}
 \prod(x,y) = \frac{A\sqrt{3}\pi}{T^{3/2}} e^{-m_0T} \int dM^2
 M^{1/2} e^{-MT}  \cong \frac{A\sqrt{3}\pi \Gamma (\frac{5}{2})
 e^{-m_0T}}{T^4}
 \end{equation}
 This is exactly the form used in the text, e.g. in Eq.(62), and in the
   Appendix 2, Eq.(A. 2.8).

\newpage

\underline{\bf APPENDIX 4}\\

{\bf Gluon spin term in the gluon propagator}\\

\setcounter{equation}{0}
\def\theequation{A4.\arabic{equation}}

The gluon propagator in the background field, Eq.(8) can be written in the
form
\begin{equation}
G=(\hat{D}^2_{\lambda}\cdot \hat{1} + ig \hat{\sum} F)^{-1}
\end{equation}
where $\hat{1}$ and $\hat{\sum}$ are  unit and spin matrices of the gluon
respectively
\begin{equation}
ig (\sum F)_{\mu\nu}= -2 ig F_{\mu\nu}
\end{equation}
 In the situation, when $G$ enters
in the gauge-invariant combination
with two parallel transporters as
in (26), (37), one can write the
Wilson loop average as in (37)
using the cluster expantion. E.g. for the adjoint Wilson loop
\begin{equation}
<W_{adj}> = <\hat{\Phi}_F(x,y,c_1)\hat{\Phi}_F(y,x,c_2)>
\end{equation}
where $\hat{\Phi}_F$ is defined in (36), and the paths $C_1,C_2$ form the
closed loop $C=C_1+C_2$ one obtains [32]
\begin{equation}
<W_{adj}> = exp \sum^{\infty}_{n=1} \frac{(ig)^n}{n!}\int \Delta (1)
...\Delta (n) \ll F(1)...F(n)\gg
\end{equation}

 We have used the following compact notation
 \begin{equation}
\Delta(k) = d\sigma(u_k) - \sum^{(1)}
d\tau_k+\sum^{(2)}d\bar{\tau}_k
\end{equation}
with the condition, that e.g.the combination
\begin{equation}
\sum^{(i)} d\tau_k\ll F(1)...F(k)...F(n)\gg=\sum^{(i)}d\tau_k
\ll F(1)...F(z(\tau_k))...F(n)\gg
\end{equation}
where $z(\tau_k)$ lies on the path $C_i$. At the  same time the combination
$d\sigma(u_k)$ is multiplied with $F(u_k)$ where $u_k$ lies on the surface $
S_{min}$ inside the contour $C$.
Therefore keeping in $\Delta (k)$
only  the first term $d\sigma(u_k)$
one obtains the usual area law
at large distances [13]:
\begin{equation}
<W_{adj}>_{no ~spin}= exp(-\sigma_{adj}S_{min})
\end{equation}

In contrast  to that the terms with $\sum^{(1)}$ or $\sum^{(2)}$ generate
perimeter-type contribution.
Indeed the cumulants $\ll F(1)...F(n)\gg$ are effectively nonzero when all
coordinates are within the interval of the size $T_g$, vacuum correlation
length, $T_g\approx 0.2 fm$ [31].

The operator $\sum^{(i)}d\tau_k$ enforces, as we discussed above, one of the
coordinates of the cumulant to lie on the contour and therefore all others
are within the distance $ T_g$ from the contour, leading to the
perimeter- type contribution
\begin{equation}
<W_{adj}>_{spin} = exp(-\sigma_{adj}S_{min}-bL)
\end{equation}
where $b$ is some constant and $L$ -- the length of the contour $C$.

For large contours $C$, which are studied in this paper, one can therefore
neglect in the first approximation the gluon spin contribution as compared
to the area-law term.
\newpage

\underline{\bf APPENDIX 5}\\

{\bf IR renormalons and background fields}\\

\setcounter{equation}{0}
\def\theequation{A5.\arabic{equation}}

The integral in (68) can be rewritten as
\begin{equation}
I\equiv \int ^{Q^2}_0 \frac{k^2dk^2}{Q^4}\ln^n
\frac{Q^2+m^2}{k^2+m^2}= \int^{n_0}_0
t^ne^{-2t}dt-\frac{m^2}{Q^2}\int^{n_0}_0t^n e^{-t}dt
\end{equation}
where $n_0=\ln\frac{Q^2+m^2}{m^2}$.

The sum in (68) can be now calculated easily (the interchange of the order
of summing and integration is possible because the integral is in finite
limits, and the sum is converging uniformly for $t\preceq n_0$). One has
\begin{equation}
\sum^{\infty}_{n=1}(\alpha_s(Q^2)t\frac{b_0}{4\pi})^n=\frac{\alpha_s(Q^2)t\cdot
b_0/4\pi}{1-\alpha_s(Q^2)
\frac{b_0}{4\pi}t}
\end{equation}
and the contribution of $I$ in the sum over $n$ is
\begin{equation}
\frac{b_0}{4\pi}
\alpha_s
(Q^2)\int^{n_0}_0
\frac{e^{-2t}tdt}{1-\alpha_s(Q^2)\frac{b_0}{4\pi}t}=\int^{n_0}_0\frac
{tdte^{-2t}}{t_p-t} \sim \frac{1}{t_p}
\end{equation}
where
\begin{equation}
t_p=\frac{4\pi}{b_0\alpha_s
(Q^2)} =\ln \frac{Q^2+m^2}{\Lambda^2}>n_0~~,
\end{equation}
since $m>\Lambda$.

Insertion of (A 5.3) into (68)  finally yields
\begin{equation}
\Delta \prod = \frac{\alpha_s(Q^2)}{8\pi^3}\cdot\frac{1}{t_p}
\cong \frac{b_0\alpha_s^2(Q^2)}{32\pi^4}
\end{equation}
This result is used in Eq.(73) of the main text.

\newpage
{\bf Figure captions}\\
\vspace{1cm}

Fig.1. The rectangular Wilson loop of the size $R\times T$ used to define
the static quark energy $E(R)$

Fig. 2. The gluon loop with external lines corresponding to the background
field $B_{\mu}$, used in [11] to calculate $\beta$ function.

Fig. 3. The same as in Fig.2 but for the ghost loop.

Fig. 4. The nonperturbative generalization of the diagrams of Fig.2 and 3
with the full account of the background in the propagators of the gluon (a)
and the ghost (b).

Fig. 5. One-gluon-exchange in the Wilson loop between the points $x$ and
$y$. Indices $\alpha, \delta$ and $\beta, \gamma$ are the color fundamental
indices of the gluon at points $x$ and $y$ respectively.

Fig. 6. The gluon-loop diagram for the gluon exchange inside Wilson
loop.

Fig. 7. The same for the ghost-loop diagram.

Fig. 8. Two-gluon exchange diagram from iteration of one-gluon-exchange
interaction.

Fig. 9. Two-gluon exchange  crossed diagram, containing both iteration of
one-gluon exchange (abelian part) and the nonabelian part.

Fig.10. Gluon exchange with the three-gluon vertex in the Wilson loop.

Fig.11. At large $N_c$ the one-gluon-exchange diagram of Fig. 5 goes over
into the product of two Wilson loops.

 Fig.12. Multiple iterative gluon exchanges.

 Fig.13. At large $N_c$ the  diagrams of Fig.6 and 7 contain  three
 Wilson loops with areas $S_1,S_2,S_3$.

 Fig.14. At large $N_c$ the  diagram of Fig.10 contains three
 Wilson loops with areas $S_1,S_2,S_3$.

Fig.15. The diagrams of Fig.4 contain at large $N_c$
 three Wilson loops.

 Fig.16. Cutting the diagram of Fig.15 along the dash-dotted line reveals
 two gluons 1 and 2, connected by two fundamental strings. One of the
 strings passes necessarily through the point 3, lying on the parallel
 transporter $\Phi(B)$, shown in Fig. 4 and Fig. 15.

 Fig.17. Diagrams containing the renormalon contribution to the correlator
 of e.m. currents.

 \end{document}